\documentclass{ws-mpla}
\usepackage[super]{cite}
\usepackage{graphicx}
\usepackage{xcolor}
\begin{document}

\markboth{A. Patk\'os}
{Isospin splitting and axion mass}

\catchline{}{}{}{}{}

\title{Isospin symmetry breaking and the mass of the QCD axion\\ in a three-flavor linear sigma model}

\author{A. PATK\'OS}

\address{Institute of Physics, E\"otv\"os Lor\'and University, P\'azm\'any P\'eter s\'et\'any 1/A\\
Budapest, H-1117, Hungary\\
patkos@galaxy.elte.hu
}

\maketitle

\pub{Received (Day Month Year)}{Revised (Day Month Year)}

\begin{abstract}
The topology of the ground state of QCD manifests itself through the dependence of its potential energy density on the CP-violating angle $\Theta$. This is computed in the present note within a three-flavor linear meson model. It is shown that the isospin symmetry breaking condensate leads to a 5\% shift in the scale of topological fluctuations, due to which a value emerges in close agreement with lattice QCD determinations. Transparent analytical computation  makes explicit the physical content of each step significantly correcting the starting crude estimate of topological susceptibility and eventually achieving the quite accurate  final result. This clear physical interpretation lends also a certain pedagogical value to the proposed treatment.

\end{abstract}

\ccode{PACS Nos.: 11.30Qc,11.30Rd,14.80Mz}

\section{Introduction}
Linear meson models have played important role from the very start of the exploration of the nature and  consequences of the $U_A(1)$ anomaly of QCD \cite{veneziano79,divecchia80,witten80}. In those models also a pseudoscalar field representing the anomalous $U_A(1)$ charge density was introduced in addition to the relevant pseudoscalar mesons. The corresponding scale was either fixed by some data of the $\eta$-sector \cite{veneziano79} or identified externally with the topological fluctuation scale of the pure gluon theory \cite{luciano18}: $\chi_{gluon,top}^{1/4}\approx 180\textrm{MeV}$. A proposal to fix the corresponding free parameter by requiring agreement with the best known estimates of the topological fluctuation scale was also made \cite{fejos25}. Another way to represent the influence of topologically non-trivial gauge configurations on meson dynamics  has been proposed  by 't Hooft \cite{tHooft76,tHooft86}
(see also \cite{rosenzweig80}). With his completion of the $U(N_f)$ invariant meson Lagrangian  extended investigations of the scalar and pseudoscalar mesons spectra have been conducted mostly concentrating on finite temperature transitions \cite{schaefer94,lenaghan00,herpay05,fejos22,tiwari23}. Possible role of operators corresponding to topologically higher charged configurations was recently also emphasized \cite{pisarski24} (see also \cite{kovacs24,fejos24}).  The parameters of these models are fixed exclusively with help of observed meson spectroscopic data. A first fully perturbative attempt to compute the two-point function of the topological charge density in the linear meson theory \cite{jiang16} led to $\chi_{top}^{1/4}\approx 165 \textrm{MeV}$.  More recent determinations were based on the dependence of the potential energy density on the angle $\Theta$ of the axial $U_A(1)$ transformations \cite{luciano18, bottaro20}. 

Similar strategy was followed by exploiting the $\Theta$-dependence of the quark mass parameters in the effective non-linear chiral models inherited from transforming away the potentially $CP$ violating piece of the pure gluonic theory  \cite{gasser82,dicortona16}. This strategy is also applicable to the effective Nambu--Jona-Lasinio model of interacting quarks \cite{linlin24}, realized both for $N_f=2,3$ cases.  

Improved realisations of all non-perturbative strategies reached the level where they find $\chi_{top}^{1/4}\in (75-77)\textrm{MeV}$, in considerable agreement with results of exact lattice simulations of QCD \cite{borsanyi16}.

In this short note the computations in the three-flavor linear meson model will be performed in well-separated steps which allows to disentangle the mechanism of reducing the starting characteristic topological scale $\sim160-180\textrm{MeV}$, first to $\sim 80\textrm{MeV}$ and eventually to a value lying in the close neighbourhood of the best actual estimates. By the well-known relationship of $\chi_{top}$ and the mass of the hypothetical axion particle \cite{weinberg78}  our analysis reveals also from this point of view  the importance of carefully taking into account the violation of isospin symmetry of strong interactions. The accurate separation of strong and electromagnetic effects in the mass splitting between the charged and neutral particles of any given iso-multiplet influences decisively the mass prediction of the QCD axion. 

In the first part of our note the isospin violating piece of the chiral condensate will be quantitatively characterized. 
The basic observation on which the whole short analysis relies is that isospin breaking effects can be calculated perturbatively on the top of any isospin-symmetric solution of the effective model. Our chosen non-perturbative solution is the Local Potential Approximation to the isospin-symmetric equations of Functional Renormalisation Group, as presented in detail with increasing accuracy in our earlier investigations. In a sense the present publication can be considered as a complement to Ref.\refcite{fejos22}.  It should be emphasized that the parameters of the effective model we shall utilize were determined without any reference to the $\eta-\eta^\prime-\pi^0$ sector which is investigated in the present note. 

In the second part the isospin violating effect will be shown to play an important role in pushing the topological susceptibility prediction into the commonly accepted range. Our calculational technique is quite common and its steps can be designed in a  straightforward manner. Still, the present aspect of the role of isospin violation was not yet discussed in quantitative detail. Isospin breaking was studied recently at the mean field level within the framework of an extended linear sigma model \cite{kovacs24}, where $U_A(1)$ anomaly was treated with a dimension-6 operator in place of the original $S_{'tHooft}$. Another version extended with coupling to the Polyakov loop was studied \cite{tawfik19} at finite isospin asymmetry applying nonzero isospin chemical potential.
Therefore the present discussion has some novel features beyond its pedagogical interest.

\section{Three-flavor linear sigma model with isospin symmetry violation}

The effective dynamics of the complex meson matrix $M$ is defined as
\begin{equation}
S_M=S_{U(3)}+S_{'tHooft}
\label{action}
\end{equation}
with
\begin{equation}
S_{U(3)}=\int_x\Big[\textrm{Tr}(\partial_{i}M^\dagger\partial_{i}M)+U(\rho)+C\tau-\textrm{Tr}[H(M+M^\dagger)]\Big], 
\label{U3action}
\end{equation}
depending on the quadratic and quartic invariants $\rho=\textrm{Tr}(M^\dagger M),\tau=\textrm{Tr}(M^\dagger M-\rho/3)^2$ and using the perturbatively renormalisable potential $U(\rho)=\lambda_1\rho+\lambda_2\rho^2$.
In the second term of  (\ref{action}) the dependence on the parameter $\Theta$ reflecting its transformation under axial $U_A(1)$ transformations is explicitly displayed:  
\begin{eqnarray}
&\displaystyle
S_{'tHooft}(\Theta)=A\int d^4x(e^{-i\Theta}\textrm{det}M+e^{i\Theta}\textrm{det}M^\dagger  ) \nonumber\\
&\displaystyle
=A\int d^4x\left[\cos\Theta(\textrm{det}M+\textrm{det}M^\dagger)-i\sin\Theta
(\textrm{det}M-\textrm{det}M^\dagger)\right].
\label{thooft-action}
\end{eqnarray}
 In this first part we shall fix $\Theta=0$ which is its observed  value. The complex field variable is defined as
\begin{equation}
M(x)=(s_a(x)+i\pi_a(x))t^a,\qquad a=0,1,2,...,8
\end{equation}
with the U(3) generators $t^a$.

Symmetry breaking background is introduced with an isospin symmetry violating condensate component in addition to the strange ($s$) and nonstrange ($ns$) components:
\begin{equation}
M_0=v_{ns}t^{ns}+v_st^s+v_3t^3, \qquad t^{ns}=\frac{1}{\sqrt{3}}(t^0-\sqrt{2}t^8),\qquad t^s=\frac{1}{\sqrt{3}}(\sqrt{2}t^0+t^8).
\end{equation}
The field equations which determine the pattern of chiral symmetry breaking on the given background are the following:
\begin{eqnarray}
&\displaystyle
\frac{\delta S_M}{\delta s_{ns}}=0=v_{ns}\left[U^\prime+\frac{A}{\sqrt{2}}v_s+C\left(\frac{1}{6}v_{ns}^2+\frac{7}{6}v_3^2-\frac{1}{3}v_s^2\right)\right]-H_{ns},\nonumber\\
&\displaystyle
\frac{\delta S_M}{\delta s_{3}}=0=v_3\left[U^\prime-\frac{A}{\sqrt{2}}v_s+C\left(\frac{1}{6}v_{3}^2+\frac{7}{6}v_{ns}^2-\frac{1}{3}v_s^2\right)\right]-H_3,\nonumber\\
&\displaystyle
\frac{\delta S_M}{\delta s_{s}}=0=v_s\left[U^\prime+C\left(\frac{2}{3}v_s^2-\frac{1}{3}(v_{ns}^2+v_3^2)\right)\right]+\frac{A}{2\sqrt{2}}(v_{ns}^2-v_3^2)-H_s.
\end{eqnarray}
The second derivative tensor of the potential energy determines the masses. For us the pseudoscalar masses are of particular interest. The diagonal second derivatives with respect to $\pi_l~(l=1,2,4,5,6,7)$ do not mix with other elements and  are the following:
\begin{eqnarray}
&\displaystyle
m^2_{1,2}\equiv m^2_{\pi^\pm}=\frac{A}{\sqrt{2}}v_s+\frac{C}{2}(v_{ns}^2+3v_3^2)+\tilde V_0,\nonumber\\
&\displaystyle
 m^2_{4,5}\equiv m^2_{K^\pm}=\frac{A}{2}(v_{ns}-v_3)+\frac{C}{2}\left(2v_s^2+(v_{ns}+v_3)^2-\sqrt{2}v_s(v_{ns}+v_3)\right)+\tilde V_0\nonumber\\
&
\displaystyle
m^2_{6,7}\equiv m^2_{K^0}=\frac{A}{2}(v_{ns}+v_3)+\frac{C}{2}\left(2v_s^2+(v_{ns}-v_3)^2-\sqrt{2}v_s(v_{ns}-v_3)\right)+\tilde V_0,
\end{eqnarray}
with $\tilde V_0=U^\prime(\rho_0)-\frac{2}{3}C\rho_0, U^\prime(\rho_0)=\lambda_1+2\lambda_2\rho$. In the above equations correspondence to the physical meson masses is hinted. These masses can be fitted to their observed (isospin violating!) values with appropriately chosen external sources in view of the following reexpression of the right hand sides of the field equations:
\begin{equation}
H_{ns}=v_{ns}m_{\pi^\pm}^2,\qquad
H_s=v_sm_{K+}^2+\frac{v_{ns}}{\sqrt{2}}(m_{K+}^2-m_{\pi^\pm}^2)-\frac{v_3}{2\sqrt{2}}m_{K-}^2,
\end{equation}
with $m^2_{K+}=(m^2_{K^\pm}+m^2_{K^0})/2, m^2_{K-}=m^2_{K^0}-m^2_{K^\pm}$.
For the third field equation sustaining the isospin violating condensate $v_3$ one quickly finds
\begin{equation}H_3=v_3\left[m_{\pi^\pm}^2+Cv_{ns}^2+\sqrt{2}v_s|A|+{\cal O}(v_3^2)\right].
\end{equation}
The couplings $|A|$ and $C$ can be expressed parametrically through $m_{K+}^2-m^2_{\pi^\pm}$ and $m_{K-}^2/v_3$ (see below (\ref{v3-determination})). Substituting these expressions into the field equation one finds a combination of $v_3(m_{K+}^2-m^2_{\pi^\pm})$ and $m_{K-}^2$, the latter being itself linear in $v_3$.

The mass matrix in the $(\pi_{ns},\pi_s,\pi_3)$ space is characterized by the following non-diagonal matrix given as a sum, where exclusively the second term depends on $v_3$:
\begin{eqnarray}
&\displaystyle
{\cal M}^2={\cal M}_0^2+{\cal M}_3^2,\nonumber\\
&\displaystyle
{\cal M}_0^2=
\begin{bmatrix}
m^2_{\pi^\pm}-\sqrt{2}A\cos\Theta v_s&-\frac{A\cos\Theta}{\sqrt{2}}v_{ns}&0\\
-\frac{A\cos\Theta}{\sqrt{2}}v_{ns}&m^2_{K+}+\frac{1}{\sqrt{2}}(m^2_{K+}-m^2_{\pi^\pm})\frac{v_{ns}}{v_s}-\frac{A\cos\Theta}{2\sqrt{2}}\frac{v_{ns}^2}{v_s}&0\\
0&0&m_{\pi^\pm}^2
\end{bmatrix},\nonumber\\
&\displaystyle
{\cal M}_3^2=\begin{bmatrix}
-Cv_3^2&0&Cv_{ns}v_3\\
0&-\frac{1}{2\sqrt{2}}\frac{m^2_{K-}}{v_s}v_3&-\frac{A\cos\Theta}{\sqrt{2}}v_3\\
Cv_{ns}v_3&-\frac{A\cos\Theta}{\sqrt{2}}v_3&-Cv_3^2.
\end{bmatrix}
\label{mass-matrix}
\end{eqnarray}
 Here the $\Theta$-dependence is displayed.
These formulae coincide with those appearing in Ref.\refcite{herpay05}  when one sets $v_3=0, \Theta=0$.  It is remarkable, that the potential $U(\rho)$ (that is the couplings $\lambda_1,\lambda_2$) does not appear explicitly, since masses of all pseudoscalar mesons receive the same contribution from it. 

The first (critically important) question is the determination of the value of $v_3$, the isospin violating condensate  with help of the mass difference between the charged kaon and the neutral kaon fields. Here we follow  the path outlined in Ref.\refcite{clement92}:
\begin{equation}
m^2_{K-}=v_3[A+C(\sqrt{2}v_s-2v_{ns})].
\label{v3-determination}
\end{equation}
The way the values of $A$ and $C$ on the right hand side of this equation are determined merits a somewhat detailed explanation. First we note, that in first approximation one can neglect in the square bracket the effect of isospin violation, that is, one can neglect ${\cal O}(v_3^2)$ terms it might contain. In other words one can rely on the solution of the isospin symmetric version of the linear sigma model. A solution which in principle accounts fully for quantum corrections of the potential is offered by the functional renormalisation group equations \cite{wetterich93,morris94}. Previously we have applied the renormalisation group flow equations to the three-flavor linear sigma model in Local Potential Approximation \cite{fejos22} also with $\rho$-dependent $C(\rho),A(\rho)$ couplings. For the simpler parametrisation (\ref{U3action}) one can extract the two couplings by employing their values taken at the minimum of the full potential energy $\rho_{min}$. Using these fully renormalized couplings the apparently tree-level looking elements of (\ref{mass-matrix}) correspond to the projection of the nonperturbatively renormalised potential on (\ref{U3action}). From Figs. 1 and 2 of  Ref.\refcite{fejos22} one finds $A=-4.8 \textrm{GeV}, C\approx 50$. The vacuum expectations $v_{ns}$ and $v_s $ are both to good accuracy equal to $93 \textrm{MeV}$.  Here we once again emphasize that the solution to which these data correspond was obtained at $\Theta=0$.   

On the left hand side of (\ref{v3-determination}) first we use,  without trying to separate the strong and the electromagnetic part of the squared mass difference the physical masses $m_{K^0,\bar K^0}=497.65 \textrm{MeV}$, $m_{K^\pm}=493.68 \textrm{MeV}$ and find
\begin{equation}
v_3\approx -0.52 \textrm{MeV}.
\label{estimate-v3}
\end{equation}
The result leads to the conclusion that effects of isospin violation to be discussed below can be taken into account perturbatively. Also it validates the approximation of neglecting any $v_3$ dependence in the squared bracket on the right hand side of (\ref{v3-determination}). This conclusion remains valid also if one attempts to estimate the value of $m^2_{K-}$ resulting purely from QCD, after separating electromagnetic contributions. This effect increasing its value has been guessed with various techniques in the past decades starting with the classic paper \cite{gasser82} yielding $m_{K-}^2=6770.51\textrm{MeV}^2$. In Ref.\refcite{donoghue96} two estimates were derived: $7216.59, 6483.04\textrm{MeV}^2$. The  $v_3$ estimates corresponding to these papers are in the range $-(0.86-0.96)\textrm{MeV}$.

 The estimates published in QCD lattice studies \cite{borsanyi16,horsley16}  are somewhat lower as can be seen from the first two columns of Table 1.  In our further discussion we rely on the values of $v_3$ obtained from lattice simulations.

If one would simply neglect the mixing of the field $\pi_3$ with $\pi_s$ and $\pi_{ns}$ and associate $m^2_{33}$ with $m^2_{\pi^0}$ one would find for the difference of the charged and neutral pion mass squares due to strong interactions a value in the range $-(37-46) \textrm{MeV}^2$ of the right sign. It would give for the part of the pion mass difference $m_{\pi^0}-m_{\pi^\pm}|_S$ originating from strong interactions $-(.10-.17)\textrm{MeV}$, which is within the $1\sigma$ error range of the lattice simulation \cite{horsley16}.  This is compatible with  the statement that the mass difference of the charged and neutral pion is exclusively of electromagnetic origin.

It is interesting therefore to follow the effect of the $\pi^0-\eta-\eta^\prime$ mixing, since it produces similarly ${\cal O}(v_3^2)$ contribution to the mass of $\pi^0$.
The squared mass eigenvalues in the $(\pi_{ns},\pi_s)\rightarrow(\eta,\eta^\prime)$-sector are obtained first from diagonalisation of the matrix ${\cal M}_0^2$, neglecting $v_3$ by a rotation with appropriately chosen angle $\psi$:
\begin{equation}
\eta=\sin\psi\eta_{ns}+\cos\psi\eta_s,\qquad \eta^\prime=\cos\psi\eta_{ns}-\sin\psi\eta_s.
\end{equation}
One denotes the corresponding eigenvalues  as $m_\eta^2,m_{\eta^\prime}^2$. With the spectral data displayed above one arrives at the estimates, actually reproducing the LPA results \cite{fejos22}:
\begin{equation}
m_\eta=537.97 {\textrm {MeV}},\qquad m_{\eta^\prime}=963.44 {\textrm {MeV}}, \qquad \tan(2\psi)=-6.47.
\end{equation}

 In the corresponding eigenbasis one computes the additional contribution from the matrix ${\cal M}_3^2$ to the squared $\pi^0$ mass perturbatively by applying the standard rules of quantum mechanical perturbation theory with the result
\begin{eqnarray}
&\displaystyle
m_{\pi^0}^2=m_{33}^2+\frac{(m_{3ns}^2\cos\psi-m_{3s}^2\sin\psi)^2}{m_{\pi^\pm}^2-m_{\eta^\prime}^2}+\frac{(m_{3ns}^2\sin\psi+m_{3s}^2\cos\psi)^2}{m_{\pi^\pm}^2-m_{\eta}^2}\nonumber\\
&\displaystyle
=m_{\pi^\pm}^2-Cv_3^2+v_3^2\left(\frac{\left(Cv_{ns}\cos\psi+\frac{A}{\sqrt{2}}\sin\psi\right)^2}{m_{\pi^\pm}^2-m_{\eta^\prime}^2}+\frac{\left(Cv_{ns}\sin\psi-\frac{A}{\sqrt{2}}\cos\psi\right)^2}{m_{\pi^\pm}^2-m_{\eta}^2}\right).
\end{eqnarray}
It is clear that the contribution from the second order perturbation theory applied with ${\cal M}_3^2(\Theta=0)$ is also negative and  is similarly ${\cal O}(v_3^2)$ like $m_{33}^2-m_{\pi^\pm}^2$. The numerics leads to
\begin{equation}
m_{\pi^0}^2-m_{\pi^\pm}^2=-91.02v_3^2.
\end{equation}
It gives, using the lattice results of $v_3$, for the QCD part of the pion mass difference $m_{\pi^0}-m_{\pi^\pm}$ the values displayed in the third column of Table 1.
These values still agree with $1\sigma$ uncertainty with the estimate of Ref.\refcite{horsley16} (no measurement appears for this quantity in Ref.\refcite{borsanyi16}).
The results based on Refs. \refcite{gasser82,donoghue96} lie outside this range, though
 they still support qualitatively the statement that the mass difference in the pion sector is overwhelmingly electromagnetic. 

Below we proceed to the discussion of the effect of the here established characteristics of isospin-violation on the topological susceptibility (which is equivalent to the mass of the axion field interacting with the $(\pi^0-\eta-\eta^\prime)$ sector).

\section{Dynamical $\Theta$-dependence of the 't Hooft potential}

Here we determine the modification of  $\Theta$-dependence of $S_{'tHooft}$ due to fluctuations in the $(\eta_{ns}-\eta_s-\pi_3)$ sector. The fluctuations arise from the meson dynamics determined at $\Theta=0$. They couple to the $\Theta$-dependent part of $S_{'tHooft}$ through the second term of (\ref{thooft-action}).  The relevant part of the action density contains terms up to quadratic power in these fields and is written in condensed form with help of (\ref{mass-matrix}) as
\begin{equation}
U_M^{(2)}=\frac{A\cos\Theta}{2\sqrt{2}}v_s(v_{ns}^2-v_3^2)+V_iN_i+\frac{1}{2}N_i{\cal M}^2_{ij}(\Theta=0)N_j,
\label{quadratic-potential}
\end{equation}
using the vectors
\begin{equation}
{\bf N}=(\eta_{ns},\eta_s,\pi_3),\qquad{\bf  V}=+\frac{A\sin\Theta}{2\sqrt{2}}(2v_sv_{ns},v_{ns}^2-v_3^2,-2v_sv_3).
\end{equation}
One completes (\ref{quadratic-potential}) to a full square in preparing the functional integration over the shifted $N$-fields. The compensating term provides the modification of the classical $\Theta$-dependence of the potential on the isospin violating background.  The integration  introduces also a constant, $\Theta$-independent one-loop contribution into the potential which can be left out of further considerations.  We shall determine the potential energy with accuracy up to ${\cal O}(v_3^2)$, therefore it is sufficient to determine the potential to linear order in ${\cal M}_3^2$:
\begin{equation}
U_{'tHooft}=\frac{A\cos\Theta}{2\sqrt{2}}v_s(v_{ns}^2-v_3^2)-\frac{1}{2}{\bf V}{\cal M}^{-2}_0(0){\bf V}
+\frac{1}{2}{\bf V}\left({\cal M}_0^{-2}(0){\cal M}^2_3(0){\cal M}_0^{-2}                                                                                                                                                                                                                                               
(0)\right){\bf V}.
\label{theta-dependent-tHofft-potential}
\end{equation}
It is clear that the potential is periodic in $\Theta$ (the last two terms are $\sim\sin^2\Theta$).

The minimum of this potential is forced to the CP-symmetric point $\Theta_{eff}=0$ 
by an extra contribution $\Theta_{PQ}$ in addition to $\Theta_{QCD}$ which emerges from the expectation value of the hypothetical axion field coupled to the anomalous axial charge density. $\Theta_{PQ}\neq 0$ is due to the breakdown of the Peccei-Quinn $U(1)_{PQ}$ symmetry \cite{peccei77}.  (For a fresh pedagogical discussion of possible CP-symmetry violation in QCD see Ref.\refcite{ringwald26}.)  The resulting recipe for introducing the dynamically fluctating part of the axion field is to perform the replacement
\begin{equation}
\Theta\rightarrow\frac{a(x)}{f_a}.
\label{theta-axion-replacement}
\end{equation}
The mass of the  axion can be read off the quadratic term in the expansion of $U_{'tHooft}$ in $a/f_a$. It is clearly related to the second derivative of $U_{'tHooft}$ with respect to $\Theta$ at the origin which is the topological susceptibility of the model.

First, consider the expression of the leading ($v_3=0$) modification of the classical mass term, 
with the simplification allowed by the very close equality of the strange and non-strange chiral condensate, $v_{ns}=v_s\equiv v\approx 93\textrm{MeV}$:
\begin{equation}
m_a^{(class)2}=\frac{v^2}{f^2_a}X,\qquad X=\frac{|A|v}{2\sqrt{2}}
\label{classical-axion-mass}
\end{equation}
due to the axion mixing with the fields $\eta_{ns},\eta_s,\pi^3$. It comes from the second term of (\ref{theta-dependent-tHofft-potential}):
\begin{equation}
m^2_a=
m_a^{(class)2}\times 
\frac{m_{ss}^2m_{nsns}^2}{m_{ss}^2m_{nsns}^2+X(4m_{ss}^2+m_{nsns}^2)}.
\label{axion-mass-eta-backreaction}
\end{equation}
Here
\begin{equation}
m_{nsns}^2=m_{\pi^\pm}^2,\qquad m^2_{ss}=m^2_{K+}\left(1+\frac{1}{\sqrt{2}}\right)-\frac{1}{\sqrt{2}}m_{\pi^\pm}^2.
\end{equation}

Using the physical mass values of $\pi^\pm, K^\pm, (K^0,\bar K^0)$ and $|A|=4800\textrm{MeV}$, one finds from (\ref{classical-axion-mass})
\begin{equation}
m_a^{(class)2}f_a^2=(192.21\textrm{MeV})^4,
\end{equation}
which qualitatively reproduces the estimate of Ref.\refcite{veneziano79}.
Taking into account the back-reaction from the $\pi^0-\eta-\eta^\prime$ sector through (\ref{axion-mass-eta-backreaction}), e.g. without considering the isospin violating effect, one finds
\begin{equation}
m_a^2f_a^2=(79.73\textrm{MeV})^4.
\end{equation}
This last value is already quite close to the result of the best lattice simulation of the topological susceptibility  \cite{borsanyi16}, but still significantly different from it. It is therefore important to investigate the stability of this result against the effect of isospin symmetry violation and the contribution of the quantum fluctuations in the $\pi^0-\eta-\eta^\prime$ sector.   
All terms of the right hand side of (\ref{theta-dependent-tHofft-potential}) produce ${\cal O}(v_3^2)$ contributions to the axion mass, the second and third also through the $v_3$-dependence of ${\bf V}$. Below we write them in a structured way in order one could assess their relative weight:
\begin{eqnarray}
&\displaystyle
\Delta_3m_a^2f_a^2=-v_3^2\Biggl[X+2X^2\left(\frac{2}{m_{\pi^\pm}^2}+\frac{4X-m^2_{nsns}}{D}\right)\nonumber\\
&\displaystyle
+\frac{X^2}{D}\left(\frac{8(Cv^2m_{ss}^2+Xm_{nsns}^2)}{m_{\pi^\pm}^2}-\frac{\left(m_{nsns}^4\left(X-Cv^2\frac{1-\sqrt{2}}{2}\right)-4m_{ss}^4Cv^2\right)}{D}\right)\Biggr].
\end{eqnarray} 
where $D=m_{ss}^2m_{nsns}^2+X(4m_{ss}^2+m_{nsns}^2)$.
For the coefficients of $-v_3^2$ from the three terms in the square bracket  on the right hand side the following estimates are obtained after substituting numerical data:  
\begin{eqnarray}
&\displaystyle
X=157826\textrm{MeV}^2,\nonumber\\
&\displaystyle
2X^2\left(\frac{2}{m_{\pi^\pm}^2}+\frac{4X-m^2_{nsns}}{D}\right)=5228998\textrm{MeV}^2,\nonumber\\
&\displaystyle
\frac{X^2}{D}\left(\frac{8(Cv^2m_{ss}^2+Xm_{nsns}^2)}{m_{\pi^\pm}^2}-\frac{\left(m_{nsns}^4\left(X-Cv^2\frac{1-\sqrt{2}}{2}\right)-4m_{ss}^4Cv^2\right)}{D}\right)\nonumber\\
&\displaystyle
~~~~~~~~~~~~~~~~~~~~~~~~~~~~~~~~~~~~~~~~~~~~~~~~~~~~~~~~~~~~=6935599\textrm{MeV}^2.
\end{eqnarray}
The enhanced values of the second and third coefficients can be understood as arising from contributions to the quadratic forms of ${\bf V}$ generated through the $33$ matrix element of
 ${\cal M}^{-2}_0(0)$, which is $\sim 1/m_{\pi_\pm}^2$. 
For instance the first term in the expression in the second line is $4X^2/m_{\pi_\pm}^2$ which is approximately                                                                                                                                                                             
 32 times larger than the contribution from first (tree level) line. 
This first term accounts for about 97\% of the total contribution of this line.  The large enhancement is due to $|A|>>m_{\pi_\pm}$, that is to the fact that the 't Hooft anomaly coefficient is an order of magnitude larger than the pion mass.
Similarly, the coefficient contribution from the third line comes predominantly from the term 
$\sim 1/m_{\pi_\pm}^{2}$.
Putting the three contributions together one has the following estimate
\begin{equation}
\Delta_3m_a^2f_a^2=- v_3^2\times(12323423\textrm{MeV}^2).
\end{equation}

Using the naively estimated value of $v_3$ in (\ref{estimate-v3}) one finds that the modified value of the topological susceptibility moves towards the lattice result (see second row of the last column of Table 1).  It is quite remarkable that the correct separation of strong and electromagnetic mass contributions still leads to significant variation (see last column of Table 1). 
 Direct QCD simulation of the topological susceptibility \cite{borsanyi16} gives: $\chi_{top}^{1/4}=75.6(1.8)(0.9)\textrm{MeV}$ and the most accurate result obtained applying chiral perturbation theory to the non-linear sigma-model with $\Theta$-dependent quark mass matrix \cite{gorghetto18} leads to $\chi_{top}^{1/4}=75.46(29)\textrm{MeV}$. These numbers can be identified with our result through the simple relation appearing first in Ref.\refcite{weinberg78}.

\begin{table}[ht]
\tbl{Aspects of isospin violation in the meson sector and its influence on the topological susceptibility, obtained from different separations of strong and electromagnetic mass for kaons.\label{tab1}}
{\tabcolsep13pt\begin{tabular}{@{}ccccc@{}}
\toprule
&$m_{K-}^2$&$v_3$&$[m_{\pi^0}-m_{\pi^\pm}]_{strong}$&$(m_af_a)^{1/2}$\\
&$\text{MeV}^2$&\text{MeV}&\text{MeV}&\text{MeV}\\
\hline
\text{iso-symmetric}&0&0&0&79.73\\
\text{naive}&3936&-0.52&-0.10&78.02\\
\text{based on [29]}&5541&-0.74&-0.18&76.17\\
\text{based on [22]}&6150&-0.82&-0.22&75.27\\ 
\botrule
\end{tabular}}
\end{table}

In conclusion we point out once again that the present discussion is a corollary of an FRG-based computation of the meson spectra in the linear three-flavor meson model, since the relevant pseudoscalar part is parametrized with help of the effective action renormalized in that framework. 
Alternatively, one could interpret the whole setup as a determination of the topological features of the model at one-loop level. It would offer the opportunity to treat the dynamics of the meson fluctuations at non-zero  $\Theta$. Assuming (somewhat arbitrarily) unchanged values of the couplings and vacuum condensates one arrives at the same expression for the classical part of the topological susceptibility. Clearly one  would then also include the $\Theta$-dependent determinant of the fluctuations in the $\pi^0-\eta-\eta^\prime$sector. The corresponding quadratically divergent integral requires renormalisation, involving the introduction of an arbitrary renormalisation scale $\mu_a$. Accepting the Principle of Minimal Sensitivity \cite{stevenson22} one can choose for $\mu_a$ a value fully suppressing this contribution. Eventually, also the perturbative computation would lead to the same topological characterisation as presented in the main part of this note.

In summary, the presented computation demonstrates that although the isospin violating condensate is apparently of nearly negligible magnitude ($v_3/v\approx 8\times 10^{-3}$), it exerts a non-negligible $4-6\%$ shift in the characteristic scale  $\chi^{1/4}_{top}$ defined by the topological susceptibility. The relative impact on the axion mass is the double of this. The predicted value depends critically on the accurate substraction of  the electromagnetic contribution from the kaon mass difference (see Table 1).
  
\section*{Acknowledgements}
Valuable discussions with G. Fej\H os on many aspects of axial $U_A(1)$ symmetry breaking and informations from S. Katz on lattice simulations of isospin splitting of hadron spectra are kindly acknowledged. Research supported by the Hungarian National Research, Development, and
Innovation Fund under Project No. K143460.

\end{document}